# Enhancement of light absorption and oxygen vacancy formation in $CeO_2$ by transition metal doping: A DFT study

Zhao Liu[a*], Hongyang Ma[b], Charles C. Sorrell[b], Pramod Koshy[b] and Judy N. Hart[b]

a. Sino-French Institute of Nuclear Engineering and Technology, Sun Yat-sen University, Zhuhai 519082, China
b. School of Materials Science and Engineering, UNSW Sydney, NSW 2052, Australia

*Corresponding e-mail: liuzhao8@mail.sysu.edu.cn

**Abstract**

It has been demonstrated in previous experimental and computational work that doping $CeO_2$ with transition metals is an effective way of tuning its properties. However, each previous study on $CeO_2$ doping has been limited to a single or a few dopants. In this paper, we systematically study the formation energies, structural stability and electronic properties of $CeO_2$ doped with the entire range of the ten $3d$ transition metals using density functional theory (DFT) calculations at the hybrid level. The formation energies of oxygen vacancies, and their effects on electronic properties, were also considered. It is found that most of the $3d$ transition metal dopants can lower the band gap of $CeO_2$, with V and Co doping significantly reducing the band gap to less than 2.0 eV. Furthermore, all of the dopants can lower the formation energy of oxygen vacancies, and those with higher atomic numbers, particularly Cu and Zn, are most effective for this purpose. The electronic structures of doped $CeO_2$ compensated by oxygen vacancies show that the presence of oxygen vacancies can further lower the band gap for most of the dopants, with V-, Cr-, Fe-, Co-, Ni-, and Cu-doped $CeO_2$ all having band gaps of less than 2.0 eV. These results suggest that doping $CeO_2$ with $3d$ transition metals could enhance the photocatalytic performance under visible light and increase the oxygen vacancy concentration, and they could provide a meaningful guide for the design of $CeO_2$-based materials with improved photocatalytic and catalytic performance as well as enhanced ionic conductivity.

**Introduction**

$CeO_2$ has long been studied for many applications, such as photocatalysis [1-3], three-way catalysis (TWC) [4, 5], oxygen sensing [6, 7], and SOFCs (Solid Oxide Fuel Cells) [8, 9]. Its interest for these applications is due to its good semiconducting properties (~ 3 eV band gap) [10, 11], high redox activity [12, 13] and high oxygen ion conductivity [14]. Research in these fields has mostly been focused on improving the performance of $CeO_2$, by tuning the electronic band structure (*i.e.* to attain a lower band gap, and hence improved visible-light absorption) and concentration of oxygen vacancies (for higher ion conductivity and redox activity) [15].

In recent years, doping with *3d* transition metals (TM) has been found to be effective for tuning the properties of $CeO_2$ for various applications [16, 17]. For example, Yue *et al.* found that the visible-light absorption and photocatalytic activity of $CeO_2$ nanoparticles can be significantly improved by doping with Co and Mn [18]. Wang *et al.* found that the spectral absorption of $CeO_2$ could be extended from the ultra-violet range to the visible-light range using Fe doping [19], which is important for use of $CeO_2$ as a photocatalyst under solar radiation. Not only light absorption can be enhanced with doping, but also the redox and catalytic activity. For example, Zhang *et al.* used a simple hydrothermal method to prepare Cu-doped $CeO_2$ nanoparticles. They found that the Cu-doped nanoparticles had an increased performance for CO oxidation than undoped $CeO_2$ due to higher redox capability and oxygen vacancy concentration [20]. Meanwhile, scandium has been found to be effective dopant for increasing the oxygen vacancy concentration and thus improving the ion conductivity of $CeO_2$ [21, 22].

Investigating the effect of doping by experimental methods has been a frequent and prevalent topic, but it remains difficult to compare and unify the research conclusions due to the varying experimental conditions and different dopants and concentrations used. In contrast, theoretical modelling by density functional theory (DFT) has the ability to clarify the effects of doping in detail. For example, DFT can provide crystallographic and electronic information about doped $CeO_2$, which enabled Kehoe *et al.* [23] to explain the reasons for the enhanced reducibility of $CeO_2$ when doped with divalent metals. Pintos *et al.* [24] used DFT to study the properties of Mn-doped $CeO_2$ and found that oxygen vacancy formation is facilitated by Mn doping. Nolan *et al.* found that the ionic radius of trivalent dopants strongly influences defect formation in $CeO_2$ [25].

Although *3d* TM doping of $CeO_2$ has been investigated previously using both experimental and computational methods, there has not been a comparative study of both the structural and electronic effects of all the *3d* metals, or a comparison of substitutional and interstitial doping. A study that includes the whole *3d* TM series can allow trends to be identified. The present work, for the first time, reports a comprehensive examination by DFT of the effects of *3d* TM doping on $CeO_2$, considering the effects of doping on the stability and electronic properties as well as the oxygen vacancy formation.

**Computational Methods**

Density Functional Theory (DFT) calculations were done using the CRYSTAL code [26, 27]. The basis sets used for Ce, O and the *3d* TMs were taken from the CRYSTAL basis set library [26]. The hybrid B3LYP method was used because it has been found to accurately reproduce the experimentally-measured properties of systems involving $CeO_2$ [28]. As described in our previous work [29], tests were conducted with different amounts of Hartree-Fock (HF) exchange energy mixed with the DFT exchange-correlation energy. 12% HF energy was found to reproduce most accurately the experimentally-measured band gap and lattice parameters of bulk $CeO_2$, so this value was used in all subsequent calculations. The $CeO_2$ supercell used for the doping studies was a 3 × 3 × 3 expansion of the primitive unit cell, with a total of 81 atoms and a size of 11.51 Å in the *x*, *y*, and z directions. The reciprocal space for the supercell was sampled using a 6 × 6 × 6 Monkhorst-Pack *k*-point grid, which gives 16 independent *k*-points in the irreducible Brillouin zone. Convergence with respect to the number of *k*-points was checked.

The structures of doped $CeO_2$ are given in the Supporting Information (Figure S1). For substitutional doping, one Ce atom in the optimized 81-atom supercell was replaced with a TM atom (Supporting Information, Figure S1a). Substitutional doping with two TM atoms was done by replacing two Ce atoms, where one of the Ce atoms replaced was the same as for the one dopant case while the second TM was placed on a second-neighbor Ce site of the first TM atom (Supporting Information, Figure S1b). For interstitial doping, the TM atom was placed at the interstitial site at the center of the unit cell (Supporting Information, Figure S2a). In the 81-atom supercell, the dopant concentrations for substitutional doping with one dopant atom and substitutional doping with two dopant atoms are 1.23 at% and 2.46 at%, respectively, while the concentration for interstitial doping is 1.22 at%. After placing the dopant atoms in the supercells, the structure was fully relaxed including both the unit cell dimensions and the ion positions. The convergence criteria for the geometry optimization were set to $3 \times 10^{-4}$ Hartree/Bohr for the root-mean-square of force and $1.2 \times 10^{-3}$ Bohr for the root-mean-square of atomic displacement.

To study oxygen vacancy formation, a vacancy was created by removing an O atom adjacent to the TM atom for substitutional doping with one TM atom, and an O atom adjacent to one of the TM atoms but separated from the second TM atom for substitutional doping with two TM atoms. This was followed by a full structural relaxation of both the unit cell dimensions and the ion positions. The structures with oxygen vacancies are shown in Supporting Information, Figure S3.

The positions of the valence and conduction bands relative to the vacuum energy were determined by firstly calculating the 1s orbital energy for an O in the center of a 2D $CeO_2$ slab. The (100) surface was used for the slab, which had a thickness of 10 atomic layers (12.80 Å); convergence with respect to the slab thickness was checked. A $8 \times 8 \times 1$ Monkhorst-Pack grid was used. The energy differences between the 1s orbitals of O atoms as far as possible from the TM atom in the doped supercells and those of O atoms in the center of the $CeO_2$ slab were then calculated. Several different oxygen atoms approximately equidistant from the dopant atom in the doped supercells were tested and found to give the same results to within 0.1 eV. The valence and conduction band energies calculated for the doped supercells were then shifted by the value of this energy difference to obtain the band positions relative to vacuum. The results for the band energies of pure $CeO_2$ are in good agreement with reference values, *i.e.* the conduction band is at a potential around 0.5 V more positive than the proton reduction potential while the valence band potential is 1.4 V more negative than the water oxidation potential [30].

The formation energies, $\Delta E_f$, of the doped $CeO_2$ were calculated by:

$$\Delta E_f = (E_{TM-doped} + x\mu_{Ce}) - (E_{pristine} + x\mu_{TM}) \qquad (1)$$

where:

| | | |
|---|---|---|
| $E_{TM-doped}$ | = | Total energy of the TM-doped 81-atom supercell |
| $E_{pristine}$ | = | Total energy of the undoped 81-atom supercell |
| $\mu_{Ce}$ | = | Atomic chemical potential of Ce |
| $\mu_{TM}$ | = | Atomic chemical potential of the TM |
| $x$ | = | number of Ce atoms replaced; $x = 0$ for interstitial doping, $x = 1$ |

for substitutional doping with 1 dopant atom, $x = 2$ for substitutional doping with 2 dopant atoms.

The formation energy of an oxygen vacancy, $\Delta E_V$, in doped $CeO_2$ was calculated by:

$$\Delta E_V = E_{TM-doped-OV} + \mu_O - E_{TM-doped} \qquad (2)$$

where:

$E_{TM-doped-OV}$ = Total energy of the TM-doped 81-atom supercell with an oxygen vacancy

$\mu_O$ = Chemical potential of of O, taken as the energy of an oxygen atom in an isolated gas-phase $O_2$ molecule

The chemical potentials of Ce and the TM atoms were calculated by:

$$\mu_{Ce} = E_{CeO2} - 2\mu_O \quad (3)$$

$$\mu_{TM} = \frac{(E_{TM-oxide} - y\mu_O)}{x} \quad (4)$$

where:

$E_{TM-oxide}$ = Total energy of the TM-oxide, $TM_xO_y$

$E_{CeO2}$ = Total energy of $CeO_2$

The TM-oxides used for calculating the chemical potentials are the stable oxides under ambient conditions, *i.e.*, $Sc_2O_3$, $TiO_2$, $V_2O_5$, $Cr_2O_3$, $MnO_2$, $Fe_2O_3$, $Co_3O_4$, NiO, CuO, ZnO. This method for calculation of defect formation energies gives the formation energies under O-rich conditions and has been used in previous work [31, 32].

**Results and Discussion**

Figure 1 shows the optimized crystal structure of $CeO_2$ and the calculated electronic density of states (DOS). The partial densities of states (PDOS) indicate that the valence band (VB) is formed predominantly by O-*2p* orbitals, with an additional minor contribution from Ce-*4f* orbitals, while the conduction band (CB) is formed by Ce-*4f* orbitals overlapping with a small contribution from O-*2p* orbitals. This hybridization of Ce and O states in the VB and CB suggests that there is mixed ionic-covalent bonding character in $CeO_2$.

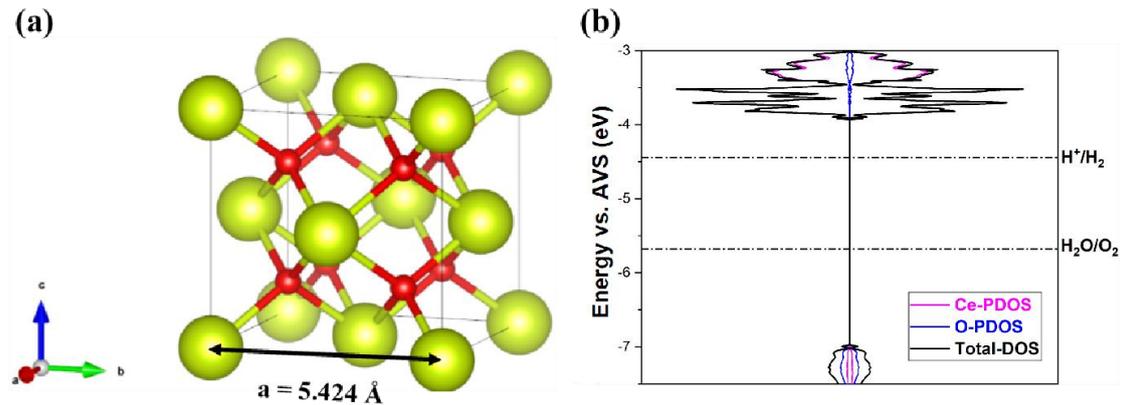

**Figure 1.** (a) Optimized crystal structure of $CeO_2$. (b) Calculated total DOS for undoped $CeO_2$, and PDOS of Ce and O, in the energy region of the band gap (energy levels are shown relative to absolute vacuum scale (AVS) for comparison with the redox potentials of the water splitting reaction, which are indicated by the horizontal dashed lines).

*Formation Energies of Doped $CeO_2$*

Figure 2a shows the calculated formation energies of $CeO_2$ doped with both one and two TM atoms at substitutional sites. The formation energy values are all positive, indicating that energy is required to achieve doping of $CeO_2$ with TMs. Formation energies for one dopant atom are in the range 1.30 - 3.85 eV, and generally increase with increasing atomic number of the dopant element, indicating decreased stability of the doped structure with increasing atomic number.

The formation energies for interstitial doping generally decrease with increasing atomic number of the dopant, but are much higher than for substitutional doping (> 4 eV), suggesting that interstitial doping with $3d$ TMs is energetically unfavorable compared with substitutional doping. Therefore, the rest of this work focuses on discussion of substitutional doping of $CeO_2$. The formation energies of interstitially-doped $CeO_2$ are shown in Supporting Information, Figure S2b.

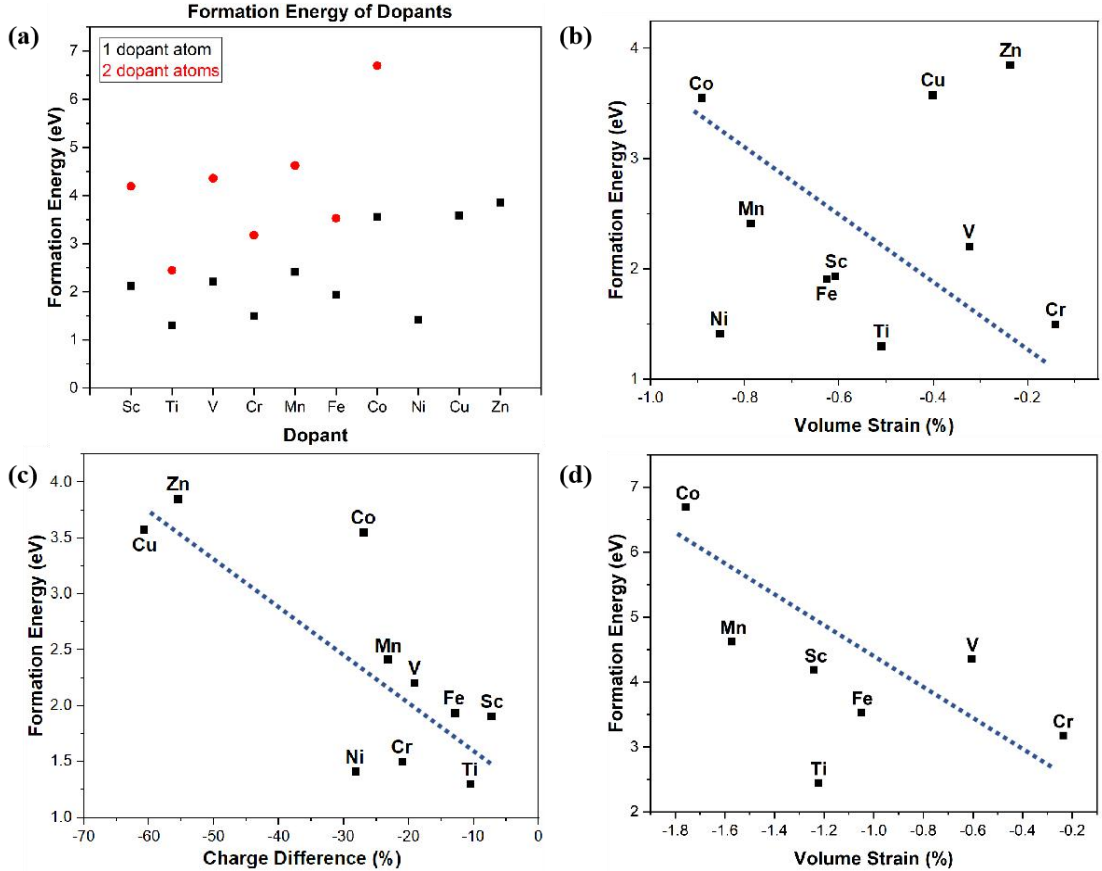

**Figure 2.** (a) Formation energies of CeO$_2$ substitutionally doped with one or two TM dopant atoms for different TM elements; (b) formation energies of CeO$_2$ doped with one TM atom as a function of volume strain[1]; (c) formation energies of CeO$_2$ doped with one TM atom as a function of charge difference[2]; (d) formation energies of CeO$_2$ doped with two TM atoms as a function of volume strain. The dotted lines are linear fits of the data and serve to guide the eye (in (b), the data points for Ni, Cu and Zn are excluded from the fit, and in (d) the data point for Ti is excluded, for reasons discussed in the text).

In general, it can be expected that formation energies of doped compounds will be predominantly affected by two factors – (1) the volume strain caused by the introduction of the dopant [33], and (2) electronic effects [34]. The relative importance of these two factors for TM-doped CeO$_2$ was analyzed. Firstly, to determine the role of volume strain, formation energies were plotted as a function of volume strain when one dopant atom was added to the $CeO_2$ supercell (Figure 2b). It can be seen that doping $CeO_2$ with *3d* TMs leads to negative volume strains, meaning that the lattice parameter of $CeO_2$ contracts upon doping, consistent with the radii of the TMs ions reported by Shannon [35] being smaller than that of $Ce^{4+}$. This result is also consistent with the experimental observation that the X-ray diffraction peaks generally shift to higher angles when $CeO_2$ is doped with *3d* TMs [36]. With the increasing magnitude of volume strain, the formation energy generally increases, indicating that volume strain does play a significant role in determining the formation energy. This is because, with more lattice distortion caused by the dopant, the structure becomes more unstable. (Although not the focus of this work, it is worth noting that interstitial doping also shows a correlation between strain magnitude and formation energy, with the strains being positive in the case of interstitial doping, Supporting Information, Figure S2b.)

---

[1] **Volume strain = (Volume of doped cell – Volume of pristine cell)/Volume of pristine cell**

[2] **Charge difference = Mulliken charge of TM – Mulliken charge of Ce in undoped CeO$_2$**

However, the formation energies for doping with Cu, Zn and Ni lie off of this general trend (and hence are excluded from the linear fit shown in Figure 2b), indicating that there are other factors that are also important in determining the formation energies for these dopants. For Cu and Zn, the formation energies are significantly higher than the general trend. When these dopants are substituted for $Ce^{4+}$ in the electrically-neutral supercell, they are effectively forced into a 4+ oxidation state [37]. Since the most stable oxidation state for Cu and Zn is 2+, these elements are expected to be highly unstable when forced into a 4+ state, resulting in relatively high formation energies. Correspondingly, the formation energy of Ti-doped $CeO_2$ lies slightly below the general trend shown in Figure 2b, consistent with Ti being most stable in a 4+ state. The possible reason for Ni lying off the general trend in Figure 2b is that it is the only dopant that has no unpaired electrons, and the spin-paired state may result in a relatively low formation energy.

The second factor that is expected to influence overall formation energies is electronic effects. To investigate these effects, the formation energies were plotted as a function of the difference between the Mulliken charges of the TM dopant atom in $CeO_2$ and Ce in undoped $CeO_2$ (Figure 2c). It can be seen that there is a clear trend of formation energy increasing as the magnitude of the charge difference increases. This suggests that a larger charge difference can cause greater instability of the doped structure. In particular, Cu and Zn fit well with this trend, and show the largest charge difference magnitude, further suggesting that for these two elements electronic effects are more important than volume strain in determining the formation energies.

In the case of adding two dopant atoms to the $CeO_2$ supercell, the formation energies are approximately two times larger than the corresponding values when a single dopant atom is added to the supercell (Figure 2a), which suggests that it becomes progressively more difficult to achieve doping as concentration increases (consistent with the formation energies being positive). Figure 2d shows the formation energies as a function of volume strain when two dopant atoms are introduced. The volume strain magnitudes are approximately double those when only a single dopant atom is added to $CeO_2$, consistent with the formation energies being approximately doubled and again suggesting that there is a strong correlation between volume strain and formation energy. The formation energy for doping with Ti lies off the general trend in Figure 2d. As discussed above, the stable charge state of Ti is 4+, so doping with Ti causes less electronic disruption to the cell and, therefore, the formation energy is relatively low.

*Formation Energies of Oxygen Vacancies in Doped $CeO_2$*
Oxygen vacancies are very important for many applications of $CeO_2$ [12, 38], so we have also studied the effect of doping on the formation of oxygen vacancies. Figure 3 shows the formation energy of an oxygen vacancy in the presence of either one or two TM atoms doped in $CeO_2$. It can be seen that, for all the TM dopants, the oxygen vacancy formation energy is lower for doped $CeO_2$ than it is for undoped $CeO_2$. This can firstly be attributed to the fact that most of the TM atoms are more stable in an oxidation state lower than 4+, and so the charge compensation provided by the oxygen vacancy allows the TM atoms to be in their more stable oxidation state. Some TM elements (e.g. Ti, V, Cr and Mn) have a similar preference for 4+ and 3+ states as Ce. For these dopants, it is likely that bond strain, i.e. weakening of the metal-oxygen bonds, is the main factor in reducing the oxygen vacancy formation energy compared with undoped $CeO_2$. Since all the 3$d$ TM ions have smaller radii than a $Ce^{4+}$ ion, when a TM atom is doped into $CeO_2$, there will be a contraction of the lattice (as discussed above and shown in Figures 2b and 2d), due to the short TM-oxygen bonds. As shown in the Supporting Information, Figure S4a, taking Ti-doped $CeO_2$ as an example, around the Ti dopant atom there are four short Ti-O bonds with an average length of 1.94 Å (compared with a Ce-O bond length of 2.35 Å in undoped $CeO_2$), while the average length of the other four Ti-O bonds is significantly increased to 2.65 Å. These significantly lengthened (i.e. weakened) bonds makes it easier to remove the oxygen that is involved in this bonding. As shown in the Supporting Information, Figure S4b-d, a similar effect is seen for doping with V, Cr and Mn, which form significantly lengthened TM-O bonds of average length 2.60 Å, 2.92 Å and 2.87 Å, respectively, owing to contraction of other TM-O bonds (Figure S4b-d). Thus, the formation energies of an oxygen vacancy are smaller for $CeO_2$ doped with these 3$d$ TM atoms than for undoped $CeO_2$.

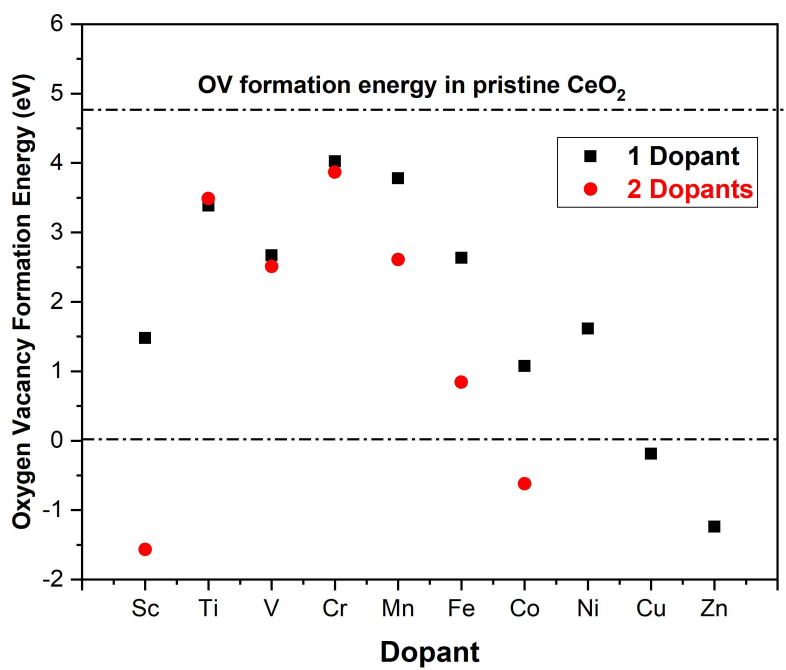

**Figure 3.** Formation energies of an oxygen vacancy in $CeO_2$ doped with one or two TM dopant atoms.

The oxygen vacancy formation energies are mostly positive, with the highest values found for those elements that can be stable in a 4+ state (Ti, V, Cr, Mn and Fe), for which charge compensation by oxygen vacancies is not required to reach a stable state. With increasing atomic number above Cr, the oxygen vacancy formation energy decreases, as the stable oxidation state moves to 3+ and then 2+; this trend is consistent with the structures without oxygen vacancies becoming more unstable with increasing atomic number, which was discussed earlier. The oxygen vacancy formation energies are negative for $CeO_2$ doped with Cu or Zn. For these two dopants, charge compensation by the oxygen vacancy balances the charge associated with the dopants being in their stable 2+ state. The fact that these two elements are very unstable in higher oxidation states makes oxygen vacancy formation highly favorable. This is consistent with the high formation energies for doping with Cu and Zn when there is no charge compensation from an oxygen vacancy (Figure 2b).

The oxygen vacancy formation energies significantly decrease as the dopant concentration increases for elements that are most stable in a 3+ oxidation state (Sc, Mn, Fe and Co), since the ratio of 2 dopant atoms to one oxygen vacancy allows this stable oxidation state to be reached. In particular, the oxygen vacancy formation energies become negative when $CeO_2$ is doped with 2 atoms of either Sc or Co, since Sc is only stable in a 3+ state while Co is mostly stable in a 3+ state. The oxygen vacancy formation energies do not significantly change with increasing dopant concentration for the other elements (Ti, V and Cr), because the ease of reduction of these elements to the 3+ state is similar to reduction of $Ce^{4+}$ to $Ce^{3+}$, so it doesn't make a significant difference if $TM^{3+}$ or $Ce^{3+}$ forms when an oxygen is removed.

*Electronic Properties*
Figure 5 shows the DOS and PDOS of pure $CeO_2$ and substitutional TM-doped $CeO_2$. It can be seen that TM doping modifies the electronic structure of $CeO_2$ and mid-gap states are introduced for most dopants. The PDOS show that the mid-gap states are predominantly formed by the dopants and they appear as narrow peaks in the DOS, suggesting that they are highly localized on the dopant atoms, due to the localized nature of the TM 3*d* orbitals [39, 40]. The unoccupied mid-gap states introduced below the conduction band generally decrease in energy with increasing atomic number, which can be attributed to the increasing magnitude of nuclear charge.

The only elements that do not introduce mid-gap states are Ti and Sc. Substituting $Ce^{4+}$ with Sc or Ti produces an empty $3d$ shell for both of these TM dopants; consistent with their relatively low atomic number, the empty TM $3d$ states are at high energies and mix with the $CeO_2$ conduction band states, so no mid-gap states are introduced (this mixing of the TM states in the conduction band can be seen in Figure 5 for Sc-doping; for Ti-doping, Figure S5 shows the DOS for a wider energy range allowing the mixing of TM states in the conduction band to be seen). For Sc-doping, the VB is partially unoccupied, due to electron transfer from the neighboring O atoms to Sc to allow the TM atom to be in its stable 3+ state. The band gap of Sc-doped $CeO_2$ is almost unchanged compared with undoped $CeO_2$. Although there are no mid-gap states in Ti-doped $CeO_2$, the valence band is shifted slightly to a more positive potential compared with undoped $CeO_2$ (Figure 5), thus doping with Ti can slightly reduce the band gap of $CeO_2$.

In contrast to doping with most of the TM elements, the mid-gap states seen for Cu- and Zn-doped $CeO_2$ are not formed by the dopants; instead they are unoccupied states formed by the nearest- and second-nearest neighbor oxygen atoms of the dopant, as shown in Figure S6. Similar to Sc-doping, this is due to electron transfer from the oxygen atoms to the TM to allow the TM to be in a lower, more stable oxidation state.

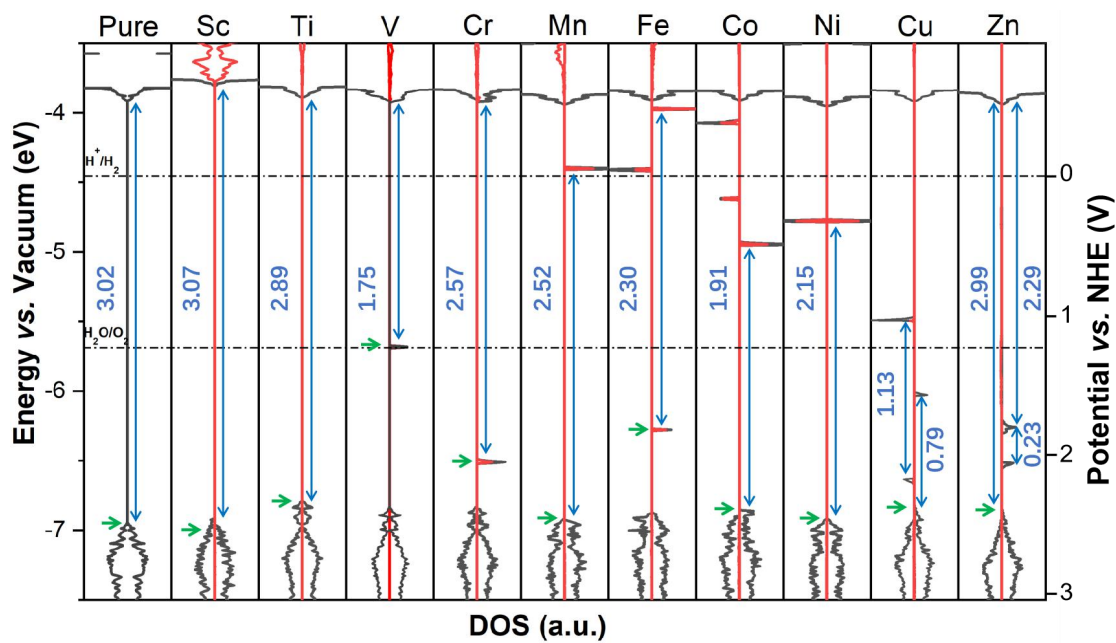

**Figure 5.** Total DOS (black lines) of TM-doped CeO$_2$; PDOS are also shown for the TM atoms (red lines). The green arrows indicate the highest occupied state for the ground state; the blue arrows and numbers indicate the lowest energy electronic transitions in eV.

Regardless of the localization, the presence of the mid-gap states can reduce the photon energy required for electrons to be excited from the highest occupied states to the lowest unoccupied states. This reduced excitation energy owing to the TM doping could improve the photocatalytic performance of CeO$_2$ under solar irradiation. With the exception of Sc-doping, the minimum energy electronic transitions are all lower than that of undoped CeO$_2$ (3.02 eV). Cu-doped CeO$_2$ has the lowest energy transition of 0.79 eV, while V- and Co-doping can also significantly reduce the minimum energy absorption of CeO$_2$ to less than 2.0 eV. In addition, the band alignments suggest that Cr-, Mn- and Fe-doped CeO$_2$ are good candidates for overall water splitting because they have decreased band gaps that straddle the oxidation and reduction potentials for O$_2$ and H$_2$ generation, respectively. In contrast, V-, Co-, Ni- and Cu-doped CeO$_2$ are only appropriate for tandem (or Z-scheme) photocatalysis [41]. Sc-, Ti- and Zn-doped CeO$_2$ have almost unchanged band gaps and band alignments compared with pristine CeO$_2$, so they should possess similar photocatalytic performance as pristine CeO$_2$. It should be noted that for Zn-doped

CeO$_2$ the transition from the valence band to the conduction band (with an energy of 2.99 eV) is considered to dominate the semiconducting behavior, although there are smaller transitions of 0.32 eV and 0.23 eV between the valence band and the unoccupied mid-gap states. Assuming that thermal energy is sufficient to excite electrons to the mid-gap states, some light absorption at an energy of 2.29 eV (corresponding to excitation from the highest mid-gap state to the conduction band) may be expected.

As discussed above, the presence of TM dopants makes oxygen vacancy formation more favorable. Oxygen vacancies can also influence electronic properties, but this influence has not been studied before. The DOS are presented in Figure 6 for CeO$_2$ doped with each of the TM elements and containing an oxygen vacancy. For each TM dopant, the DOS is shown for the structure with either one or two dopant atoms, whichever gives the lower formation energy for the oxygen vacancy. Thus, based on the vacancy formation energies shown in Figure 4, Figure 6 shows the DOS of TM-doped CeO$_2$ with one oxygen vacancy and 2 dopant atoms for doping with Sc, V, Cr, Mn, Fe and Co, and 1 dopant atom for Ti, Ni, Cu and Zn doping.

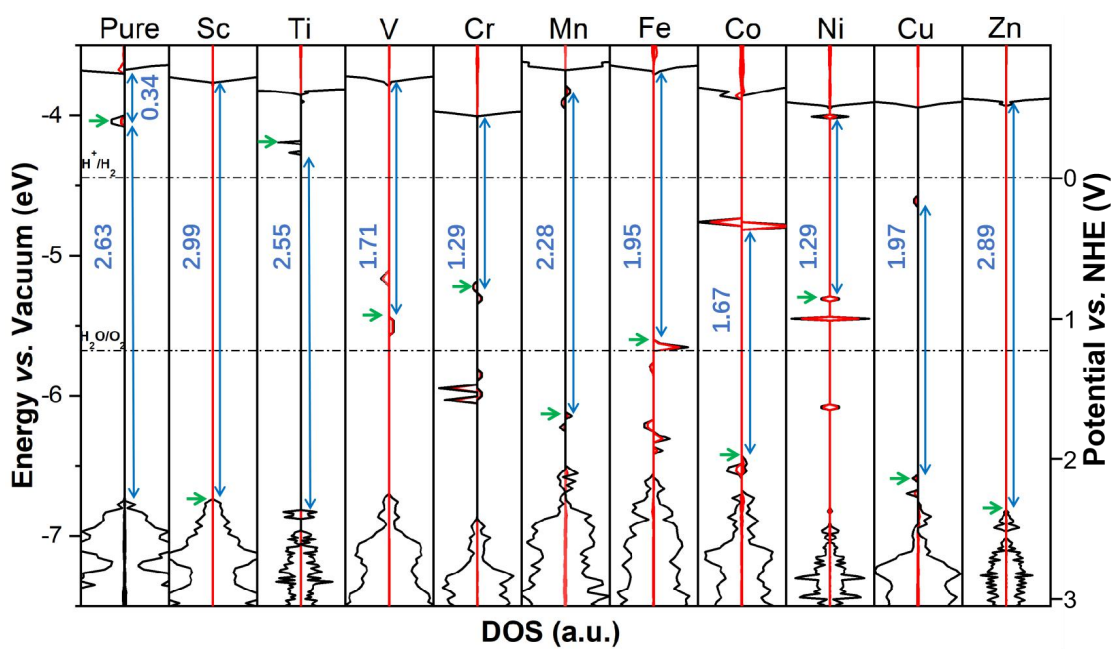

**Figure 6** Total DOS (black lines) of TM-doped CeO$_2$ with an oxygen vacancy; PDOS for the TM atoms are also shown (red lines). The DOS are shown for the dopant

concentrations that give the lowest vacancy formation energy. For Sc, V, Cr, Mn, Fe and Co, this is the structure with two dopant atoms and one vacancy. For Ti, Ni, Cu and Zn, the result is shown for the structures with one dopant atom and one vacancy. The green arrows indicate the highest occupied state for the ground state; the blue arrows and numbers indicate the lowest energy electronic transitions in eV.

In undoped $CeO_2$, when an oxygen vacancy is created, there is an occupied mid-gap state formed 0.3 eV below the conduction band; assuming that thermal energy is sufficient to promote electrons from this state to the conduction band, it results in a reduction of the minimum energy transition to 2.63 eV, compared with pristine $CeO_2$. Figure S7 shows that the mid-gap state is formed by reduced $Ce^{3+}$ near the oxygen vacancy. For the doped cases, previously unoccupied states will become occupied due to donation of electrons from the vacancy. As a result, for Ti, V, Cr, Mn, Fe, Co and Ni doping, in comparison with doping with no vacancy, some of the unoccupied states at or below the conduction bands become occupied and shift down in energy and the highest occupied states move up in energy due to these extra electrons in the 3$d$ shells. Furthermore, some of the unpaired mid-gap states become paired. For example, in 2Mn+1OV, two pairs of mid-gap states are seen, while there is only one unpaired mid-gap state when there is no vacancy (Figure 5). For 2Co+1OV, mid-gap states are paired whereas they are not paired without the vacancy present.

For 1Ti+1OV, compared with Ti doping without a vacancy, the presence of an oxygen vacancy introduces some mid-gap states, only slightly below the conduction band (by 0.33 eV). These states are formed by reduced $Ce^{3+}$ near the Ti atom. Assuming that electrons in these states can move to the conduction band due to thermal excitation, the minimum energy for an electronic transition is 2.55 eV.

For 2Sc+1OV, there is no mid-gap state, similar to the case of Sc doping with no vacancy, but the unoccupied valence band states becomes occupied when the vacancy is introduced, due to the charge compensation between the reduced Sc atoms and the oxygen vacancy. The band gap is only slightly reduced compared with undoped $CeO_2$ (to 2.99 eV). Similarly, for both 1Zn+1OV and 1Cu+1OV, the unoccupied oxygen states are removed when the vacancy is created. For 1Zn+1OV, there are no mid-gap states and the band gap does not change much compared with pristine $CeO_2$, while for 1Cu+1OV, there are mid-gap states formed by the Cu dopant.

For the dopants that do create mid-gap states, the electronic transitions are predominantly from the mid-gap states to the conduction bands (or to mid-gap states close to the conduction band). The minimum energies required for electronic transitions are reduced by 0.3-0.8 eV compared to doping with no vacancies, with the exception of Cu doping for which the minimum electronic transition energy is increased. Except for Sc, Ti, Mn and Zn doping, all the other dopants result in minimum transition energies that are less than 2.0 eV. Among the ten dopants, Cr- and Ni-doped $CeO_2$ have the lowest transition energies of 1.29 eV. However, only the four dopants that give minimum electronic transitions above 2.0 eV have band alignments suitable for overall water splitting.

*Applications of TM-doped $CeO_2$*

In general, the results presented here show that $3d$ TM doping may improve the performance of $CeO_2$ from two aspects. Firstly, the formation energy of oxygen vacancies is lowered compared with pristine $CeO_2$, therefore higher oxygen vacancy concentrations as well as higher ion conductivity are expected. The former is beneficial for applications in catalysis, as oxygen vacancies usually act as active sites in catalytic reactions [42]. The increased ionic conductivity is beneficial for applications of $CeO_2$ as an electrode or electrolyte material in SOFCs [43]. It is found here that, in general, TMs of higher atomic number are better for promoting oxygen vacancy formation. Due to them giving the lowest vacancy formation energies, Sc, Co, Cu and Zn are expected to be the most effective dopants for increasing the oxygen vacancy concentration [44]. In particular, Cu and Zn doping are promising in this regard, because despite the relatively high formation energy for adding the dopant to $CeO_2$ (Figure 2), they have relatively low overall formation energies for adding the dopant in combination with the oxygen vacancy. Furthermore, they should produce a 1:1 ratio of vacancies to dopant atoms, compared with 1:2 for Sc and Co.

This effect of dopants, particularly for atomic numbers above Mn, increasing the oxygen vacancy concentration is consistent with previous experimental work. For example, previous studies have found that both Ni-doped $CeO_2$ and Co-doped $CeO_2$ possess increased catalytic performance compared with undoped $CeO_2$ due to the higher concentration of oxygen vacancies [45, 46], while Cu-doped $CeO_2$ has also been found to have improved redox activity towards CO conversion [20, 47, 48]. Mn-doped $CeO_2$ has been found to have higher catalytic activity than undoped $CeO_2$ due, in part, to higher oxygen mobility [49], and as mentioned in the Introduction, doping $CeO_2$ with scandium has been found to improve the ion conductivity [21, 22].

The second aspect of performance that may be improved by TM doping is visible-light absorption. The DOS analysis presented here shows that $3d$ TM doping can lower the energy at which $CeO_2$ can absorb light by introducing mid-gap states, thus potentially enhancing the visible-light activity, which is beneficial for use of $CeO_2$ as a photocatalyst. Only doping with Sc, Ti and Zn is likely to be ineffective for this purpose. However, the results presented here indicate that it is important to consider the combined effect of the TM dopant and oxygen vacancies on the electronic properties, as oxygen vacancy formation may be favorable and the predicted electronic properties are very different when oxygen vacancies are present compared with only the TM dopant being present.

Again, the results presented here for the effect of dopants on the electronic properties are consistent with previous experimental studies. For example, previous experimental work has shown that Cr-doped $CeO_2$, Fe-doped $CeO_2$, Mn-doped $CeO_2$ and Co-doped $CeO_2$ all possess enhanced visible-light photocatalytic performance compared with undoped $CeO_2$ [18, 19, 50, 51], while Mn doping of $CeO_2$ has been found to systematically decrease the band gap down to 2.83 eV with increasing Mn doping concentration [52]. In contrast, Zn-doping has been found to have only a small effect on the band gap of $CeO_2$ and can in fact suppress the photocatalytic activity compared with undoped $CeO_2$ [53].

The data presented in this paper can be used as a predictive tool for design of doped $CeO_2$ to be used in various practical applications. For example, for applications in SOFCs and catalysis, where redox activity and ionic conductivity are most important, dopants that can lower the oxygen vacancy formation energy the most are ideal, which are predicted to be Sc, Co, Cu and Zn according to Figure 4. For applications, such as photocatalysis, it is important to enhance visible-light absorption. When oxygen vacancies are present (i.e. under oxygen-poor conditions, or for dopants that are likely to spontaneously form vacancies even under oxygen-rich conditions), V, Cr, Co and Ni are the best choices for dopants to reduce the minimum energy for photon absorption according to Figure 6, while Mn doping may be a good option for achieving reduced band gap while still maintaining appropriate band energies for overall water splitting. For doping under oxygen-rich conditions in cases where oxygen vacancies are not likely to be present, V is expected to be a good dopant for reducing the energy of light absorption to less than 2.0 eV, while doping with Cr, Mn or Fe may be best for overall water splitting with visible-light activity, according to Figure 5.

**Conclusions**

In this work, we have systematically studied $CeO_2$ doping by all the ten $3d$ TM elements by means of first-principles calculations. Formation energy calculations for adding the dopants suggest that it is more difficult to dope with increasing dopant concentration, and the dopants that cause smaller structural and electronic distortion of the $CeO_2$ structure are easier to incorporate. Formation energies of oxygen vacancies suggest that all the ten $3d$ TMs can increase the vacancy concentration compared with undoped $CeO_2$ by decreasing the oxygen vacancy formation energy, which is consistent with previous experimental findings that $3d$ TM doped $CeO_2$ generally has enhanced catalytic performance and ion conductivity due to the higher oxygen vacancy concentration. According to our results, Sc, Co, Cu and Zn can lower the formation energy of oxygen vacancy the most, and thus are good choices as dopants for increasing the oxygen vacancy concentration and ion conductivity. Electronic property analysis suggests that most of the TMs can lower the energy at which light can be absorbed by $CeO_2$ by introducing mid-gap states, which is also in accordance with previous experimental findings that doped $CeO_2$ often has better optical absorbance and photocatalytic performance than undoped $CeO_2$. Under oxygen-rich conditions when vacancies are not present, V may be the best choice for giving the minimum energy for photon absorption, while under conditions where oxygen vacancies form, V, Cr, Co and Ni are good choices. The present work not only provides a theoretical basis to explain previous research on the effects $CeO_2$ doping, but also provides strategies for fabrication of $CeO_2$ with better properties for applications as photocatalysts, catalysts and ion conductors.

**Acknowledgements**

The authors acknowledge financial support from the Australian Research Council (ARC). This work was supported by computational resources provided by the Australian Government through the National Computational Infrastructure (NCI) under the National Computational Merit Allocation Scheme. This work was also supported by the National Natural Science Foundation of China (NSFC 12002402, 11832019, 11472313, 13572355), the Fundamental Research Funds for the Central Universities (45000-31610046), the Guangdong Major Project of Basic and Applied Basic Research (2019B030302011), and the Guangdong Basic and Applied Basic Research Foundation (2019A050510022). Z. L. also acknowledges financial support from the Guangdong Province - Overseas Young Postdoc Recruitment Program.

**Conflict of Interest**

The authors declare no conflict of interest.

-